\documentclass[11pt,twoside]{article}
\usepackage{asp2010}

\resetcounters

\begin{document}

\title{Investigation of magnetically sensitive FeH lines}
\author{Wende,~S.$^1$, Reiners,~A.$^1$, Seifahrt,~A.$^2$, Shulyak,~D.$^1$,
 and Kochukhov,~O.$^3$
\affil{$^1$Institut f\"ur Astrophysik, Georg-August-Universit\"at G\"ottingen, Friedrich-Hund-Platz 1, D-37077 G\"ottingen, Germany}
\affil{$^2$Physics Departement Univ. of California, One Shields Avenue Davis, CA 95616 USA}
\affil{$^3$Department of Physics and Astronomy, Uppsala University, Box 516, Uppsala 751\,20, Sweden}}

\begin{abstract}
M-type stars exhibit strong magnetic fields towards decreasing effective temperatures. 
The measurement of these fields is complicated due to missing
indicators. Molecular FeH lines provide an excellent means to
determine magnetic field strengths from the Zeeman broadening of
magnetically sensitive lines.\\  
Our aim is the investigation of possible dependencies of the amount of
sensitivity to magnetic fields from rotational quantum number, branch,
and the projection of the total angular momentum onto the internuclear axis 
($\Omega$).
We also compare results from computations with those from
observations.\\
We use high resolution CRIRES spectra of the two M dwarfs GJ1002
(M5.5 inactive) and GJ1224 (M4.5 active). Individual lines are
fitted by Gaussians and the obtained line depths and widths from the
active and inactive star can be compared with each other. In this
way, magnetically sensitive lines can be detected. For test purposes,
we do the same with computed spectra of FeH. One with zero magnetic
field and the other with a $2$\,kG magnetic field vector used 
in the disc integration (i.e. pure radial at the disc center).\\
We found, in agreement with theory, that lines with high $\Omega$ show
strong sensitivity to magnetic fields. No obvious correlation with
branch or $J$ was found, which was also expected for lines formed in
intermediate Hund's case. The computations agreed in general well with
the observations, but in many cases the individual splitting of
certain lines can be very different to observations.\\

\end{abstract}

\section{Introduction}
M-type stars are most numerous in the universe and are also dominating the
stellar mass function. Towards later spectral types, they become fully
convective and exhibit strong magnetic fields: 
while only $0.8$\,\% of M0 dwarfs show H$_{\alpha}$ emission (which is
an indicator for magnetic activity), more than $70$\,\% of the M8
dwarfs show signs of magnetic activity \citep{2005nlds.book.....R}.
Measuring these magnetic fields in M-type stars is an interesting, but very
challenging task. The well probed atomic indicators are vanishing
towards these low temperatures, or become too strongly pressure
broadened. A possible solution is the application of molecular FeH
lines, which are very numerous and strong around $1000$\,nm. The
molecule provides magnetically sensitive and insensitive absorption
lines closely side by side, which makes it in principle possible to adjust 
stellar parameters of synthetic spectra to the insensitive lines and
then use the sensitive lines to obtain magnetic field
strength. Unfortunately the description of the molecular Zeeman effect
for FeH is still insufficient, since most FeH lines behave like they
were formed in intermediate Hund's case \citep{2002A&A...385..701B,2003A&A...412..513B}.    
Recent work was done by
\citet{2007A&A...473L...1A,2008A&A...482..387A}, who determined
 Land\'e $g$ factors by empirically fitting solar FeH
lines. The factors could then be used to compute spectra of FeH
including Zeeman splitting.
Another way to determine magnetic field strengths was demonstrated by
\citet{2006ApJ...644..497R,2007ApJ...656.1121R}. They compared
magnetically sensitive FeH lines from active M dwarfs to those from an
inactive template M dwarf.

In this work, we identify magnetically sensitive
FeH lines in high resolution CRIRES spectra of the active M4.5 dwarf
GJ1224, through comparison with the inactive M5.5 dwarf GJ1002.  
With the help of the line list from \citet{2010A&A...523A..58W}, we
assign quantum numbers and investigate the dependence of magnetic
sensitivity on rotational quantum number $J$, branch, and $\Omega$
(projection of the total angular momentum onto the internuclear axis).

\section{Identification of Sensitive Lines}
When comparing FeH in a spectrum of an M dwarf with a known strong
magnetic field to a  spectrum of an M dwarf with only weak magnetic
activity and similar spectral type (i.e. effective temperature), one
notices that certain lines of the magnetically active star are 
broader than their counterparts in the inactive star. This could, of
course, be due to different rotation velocities, but since only some lines are affected, 
the broadening must be due to the Zeeman effect which also operates
in molecular lines \citep{2002A&A...385..701B,2003A&A...412..513B}.  
\citet{2006ApJ...644..497R,2007ApJ...656.1121R} used this effect to
determine magnetic field strengths in a sample of M type dwarfs. To make stars
with different effective temperatures comparable, they used a scaling
procedure which is inspired by scaling optical depth:
\begin{eqnarray}
S(\lambda)=1-C(1-A(\lambda)^{\alpha}).
\label{magfeh}
\end{eqnarray}
In this expression, $S(\lambda)$ is the resulting scaled spectrum,
$A(\lambda)$ the normalized spectrum which will be scaled, $\alpha$
the optical depth scaling factor, which is applied to the overall
spectral range, and C is a constant controlling the maximum of
absorption due to saturation.  
To determine the magnetic field strength, they linearly interpolate between a
zero field template star and one with known magnetic field. The zero
field template star is the M dwarf GJ1002 which was already
used for the indentification of FeH lines in the z-band \citep{2010A&A...523A..58W}. 
For GJ1224 (M 4.5 dwarf), they determined a
magnetic field strength of $\sim 2.7$\,kG. 

\begin{figure}
\plotone{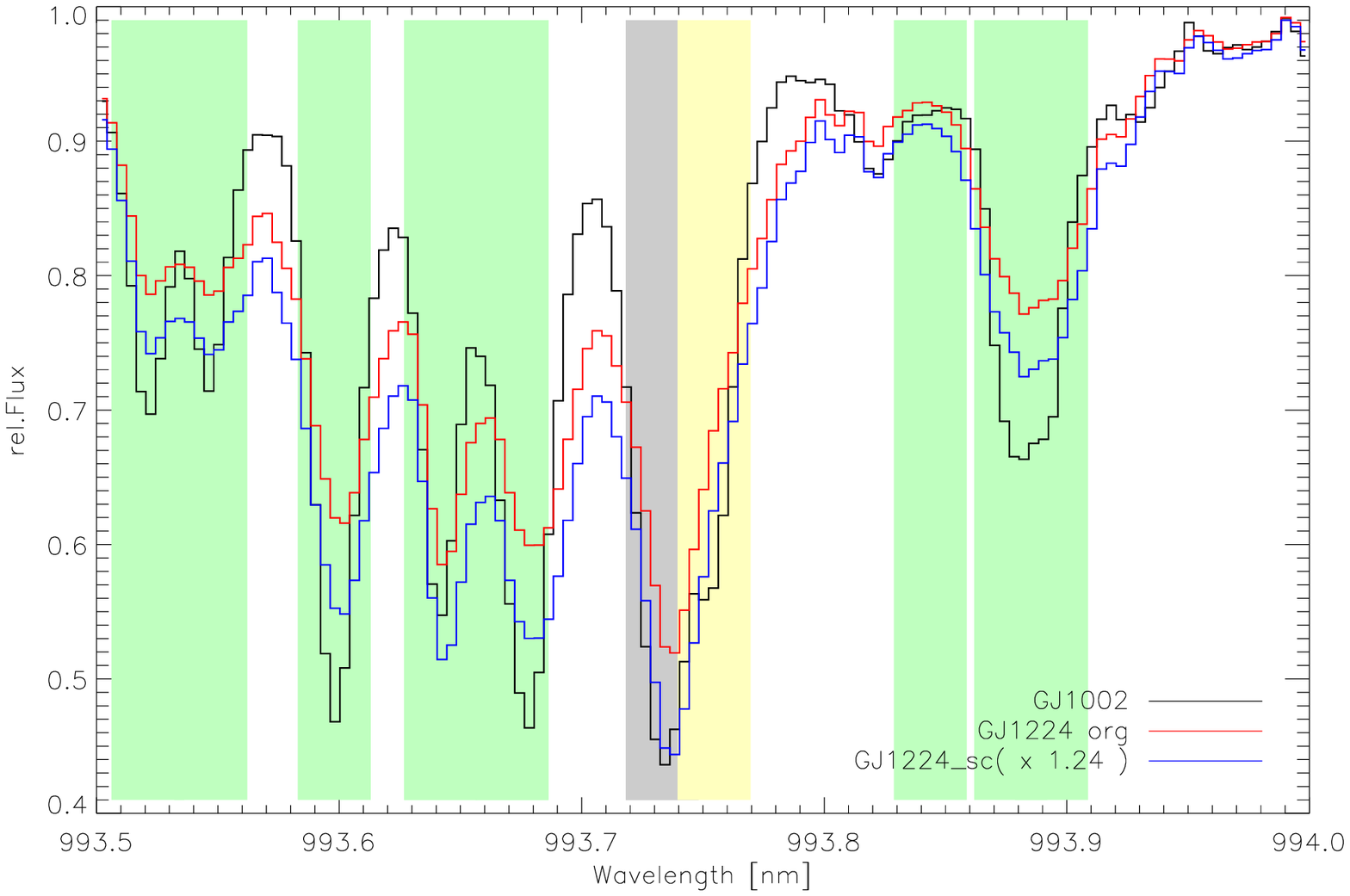}
\plotone{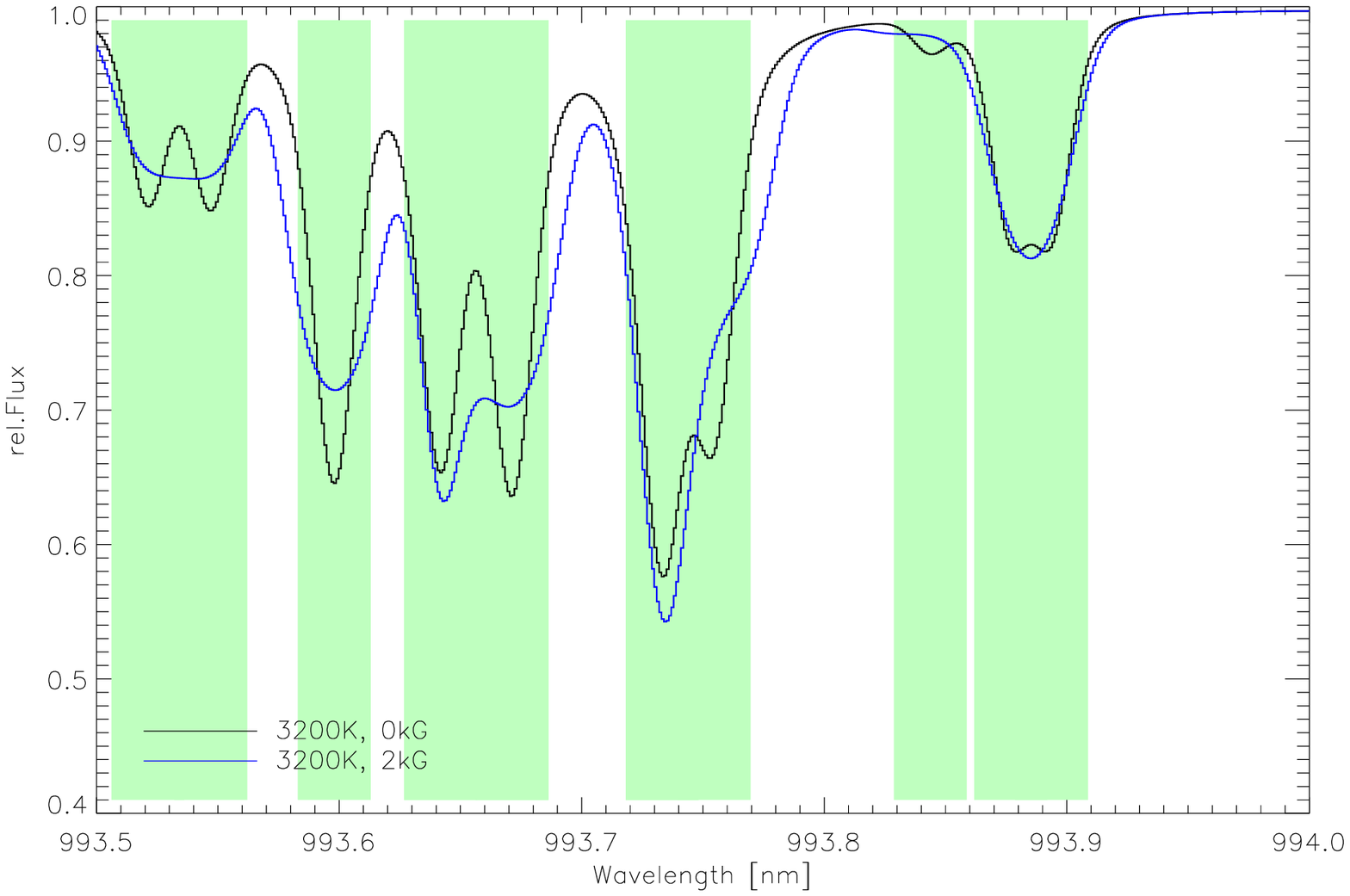}
\caption{Upper Figure: Comparison between
   GJ1224 (red unscaled and blue scaled) and GJ1002 (black). Strong
   magnetic sensitive lines are highlighted
   with green, mildly sensitive lines with yellow, and insensitive
   lines with grey. Bottom Figure: Comparison between computed spectra
   with ($2$\,kG field, blue line) and without magnetic
   field (black line).}
\label{Zeeman1}
\end{figure}

\begin{figure}
\plotone{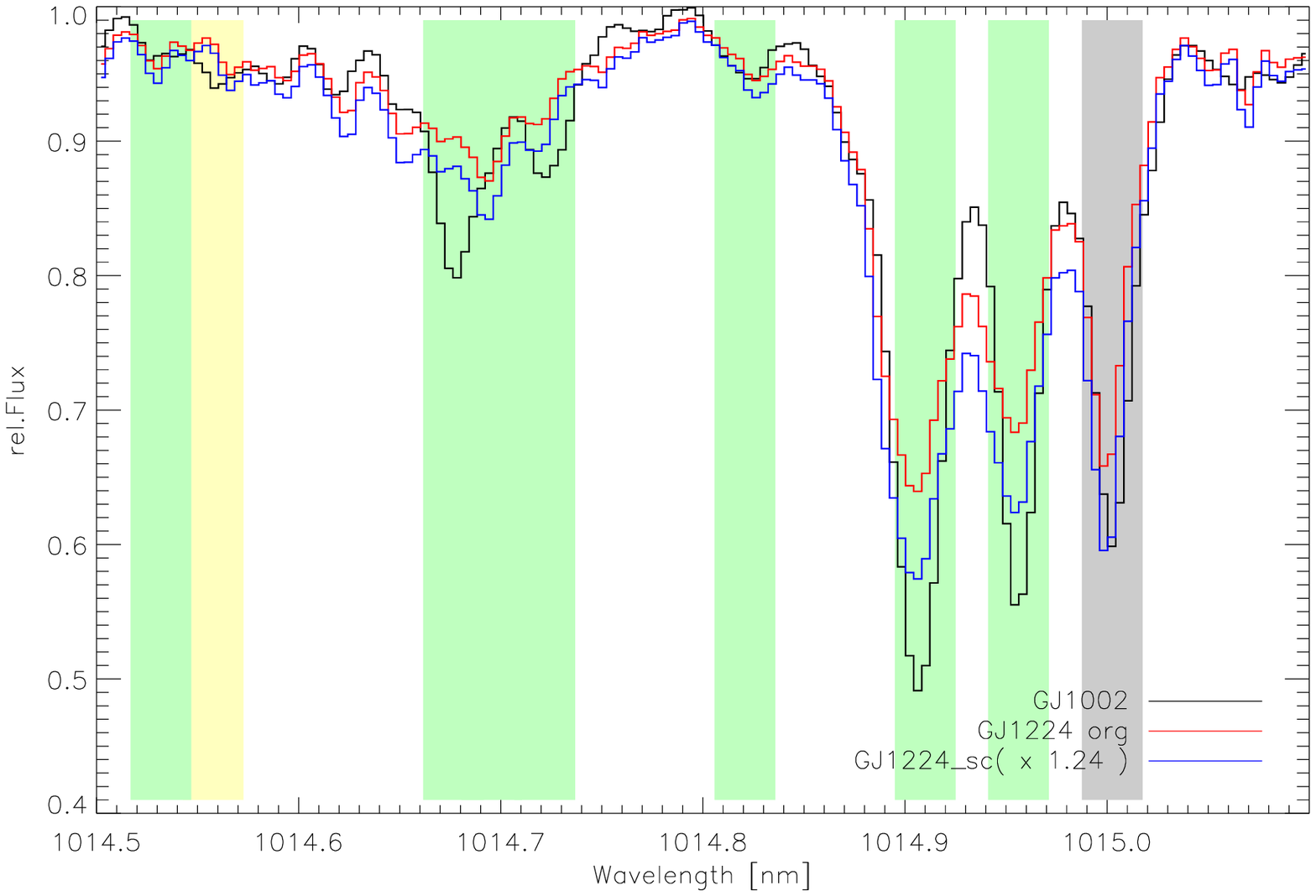}
\plotone{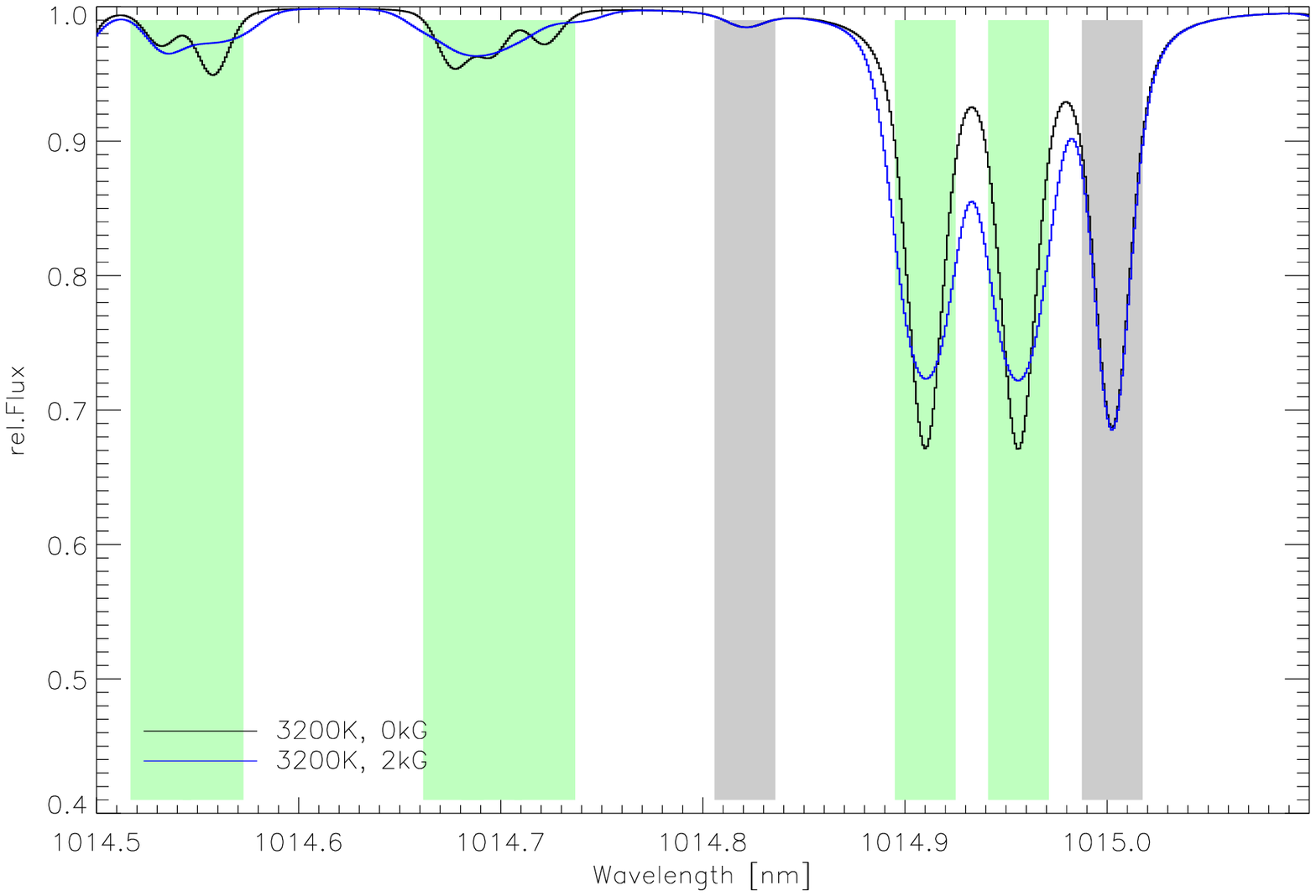}
\caption{Upper Figure: Comparison between
   GJ1224 (red unscaled and blue scaled) and GJ1002 (black). Strong
   magnetic sensitive lines are highlighted
   with green, mildly sensitive lines with yellow, and insensitive
   lines with grey. Bottom Figure: Comparison between computed spectra
   with ($2$\,kG field, blue line) and without magnetic
   field (black line).}
\label{Zeeman1b}
\end{figure}

We obtained high-resolution CRIRES\footnote{Data for GJ1224 were taken at ESO Telescopes
  under the program 83.D-0124(A). Data for GJ1002 were taken at ESO Telescopes
  under the program 79.D-0357(A).} spectra for the same stars over the whole z-range
and used it to detect more magnetically sensitive FeH lines redwards of $1\mu$\,m.
For this task, we used the optical depth scaling with $\alpha=1.24$ for
GJ1224 and compared it with the spectra of GJ1002 (both have
$v\sin{i}\le 3$\,km$^{-1}$).
Two exemplary spectral bins are shown in the upper plots of 
Figs.~\ref{Zeeman1} and ~\ref{Zeeman1b}. It is
obvious that some lines are strongly split and others not at all. The
unsplit lines were used to scale GJ1224 to the effective
temperature of GJ1002. We will quantify the identification by fitting
Gaussian line profiles to the FeH absorption lines and compare line depths and line
widths in Sect.~\ref{sec:Zeemancompare}  

\section{Theoretical Zeeman Splitting}
The theoretical description of the Zeeman effect in FeH molecular
lines is still a challenging task, since the Born-Oppenheimer approximation 
is no longer useful for determining Land\'e g factors. Also,
the rovibronic transitions
of FeH are mostly in intermediate Hund's case, and the description of
the Zeeman splitting must also be treated in this intermediate case
\citep{2002A&A...385..701B,2003A&A...412..513B}. Not all lines can
be described in this case, which make an empirical ansatz necessary
\citep{2007A&A...473L...1A,2008A&A...482..387A}. A semi-analytical
description was presented by \citet{2010A&A...523A..37S}, who found, that the
intermediate case is, in general, a good approximation for the following
cases:
\begin{enumerate}
\item $\Omega_l=0.5$
\item $\Omega_{l~or~u}\le 2.5$ and $3Y>J(J+1)$ for P and Q branches
\item $\Omega_{l~and~u}=2.5$ and $5Y>J(J+1)$ for the R branch
\end{enumerate}
 Here, $Y=|A_v/B_v|$ is the ratio of the spin-orbit coupling and
 rotational constants. For all other cases, a good approximation is the
 assumption of Hund's case (a) for the upper level and Hund's case (b)
 for the lower level. We follow this description and use a code from 
\citet{Bleroy} (modified by D. Shulyak) to
determine Land\'e factors which describe the strength of the
splitting. These factors can be
used in the \texttt{SYNMAST} code \citep{2007pms..conf..109K} to
generate spectra including effects from Zeeman splitting. 
In the bottom plots of Figs.~\ref{Zeeman1} and ~\ref{Zeeman1b}, 
the two exemplary spectral regions are shown for computed spectra
with zero magnetic
field and with a $2$\,kG magnetic field vector (it describes 
a pure radial magnetic field at the disc center and is pure horizontal 
at the limb). 
The observed and computed
spectra look similar, but at least for some lines, the computed splitting is very
different from the observed ones. These shortcomings could be related to
the inadequate theoretical description of the Zeeman splitting as well as to
possible more significant horizontal components in the geometry of the magnetic field.  
The computed spectra also show the possibility, that the line
depth could be enhanced due to the split components. That means, that
it is necessary to investigate the line width as well as the line
depth to detect magnetically sensitive lines.
\section{Comparison Between Computations and Observations }
\label{sec:Zeemancompare}
\begin{figure}
\plotone{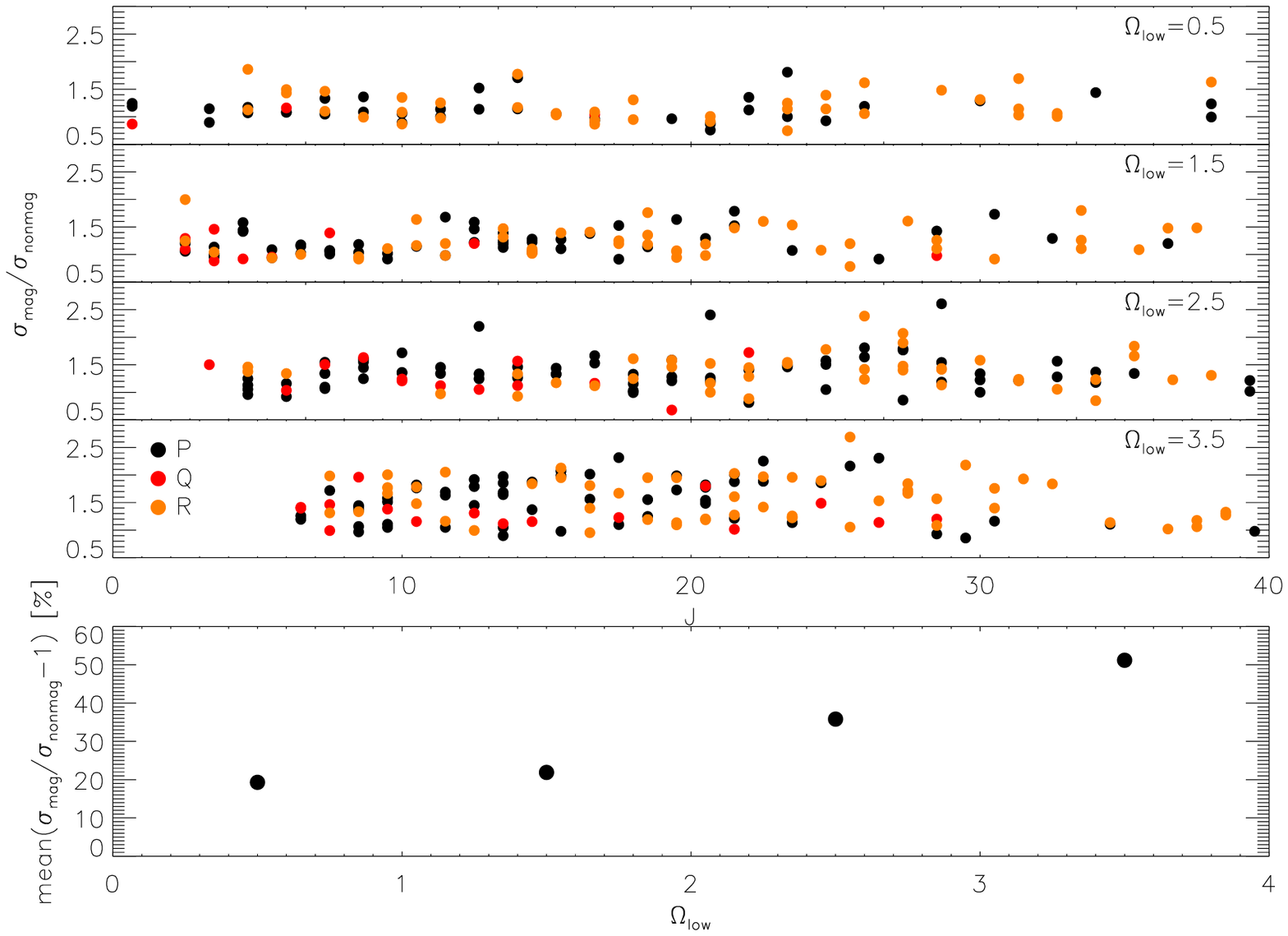}
\plotone{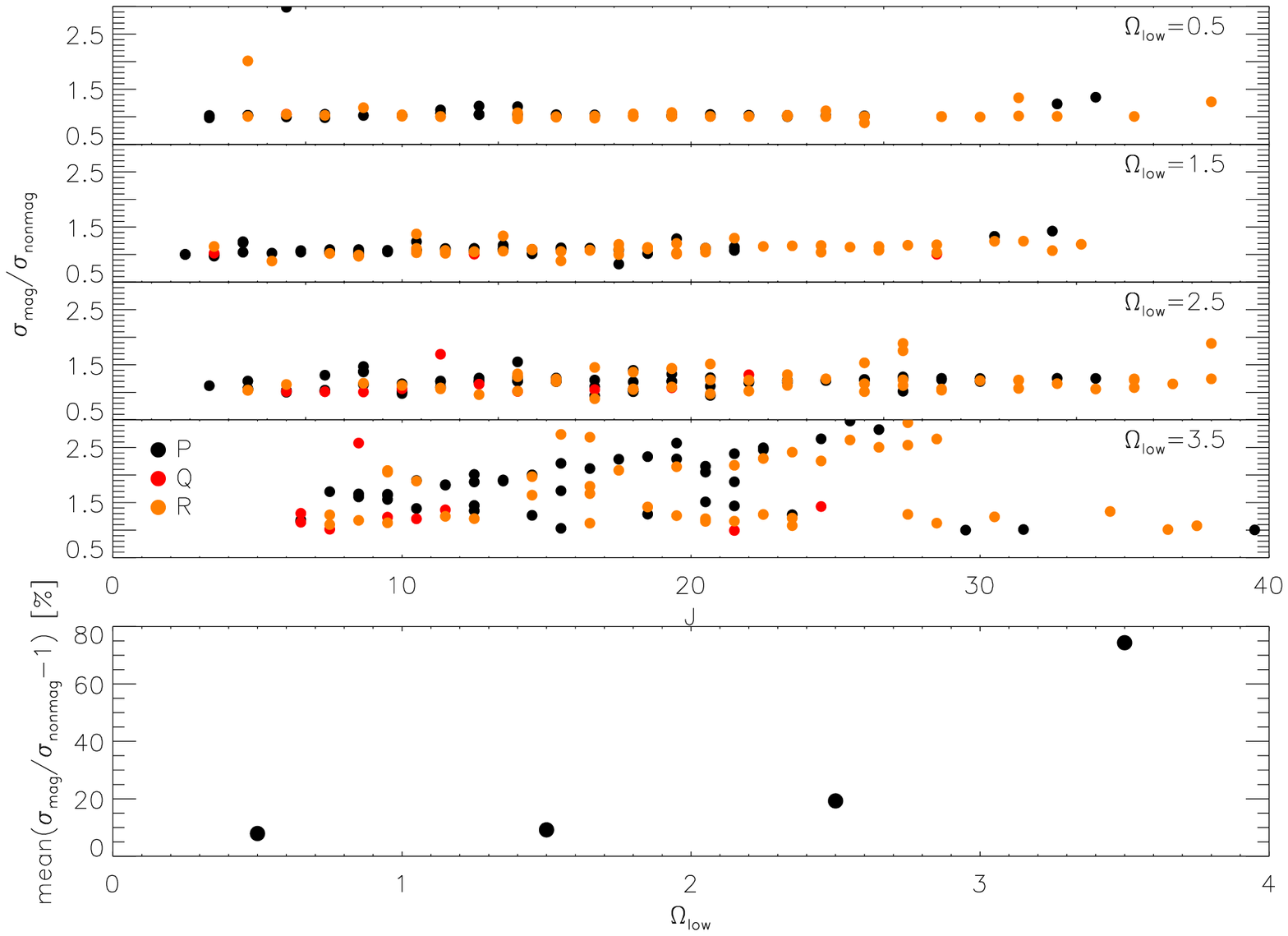}
 \caption[]{Ratio between the widths of the FeH lines in the magnetic
   and non magnetic case as a function of rotational quantum number $J$. Upper plot shows the results from the
   observations, bottom plot from the computations. The lower panels
   each show the average ratio for each $\Omega$.}
\label{Zeeman2}
\end{figure}

\begin{figure}
\plotone{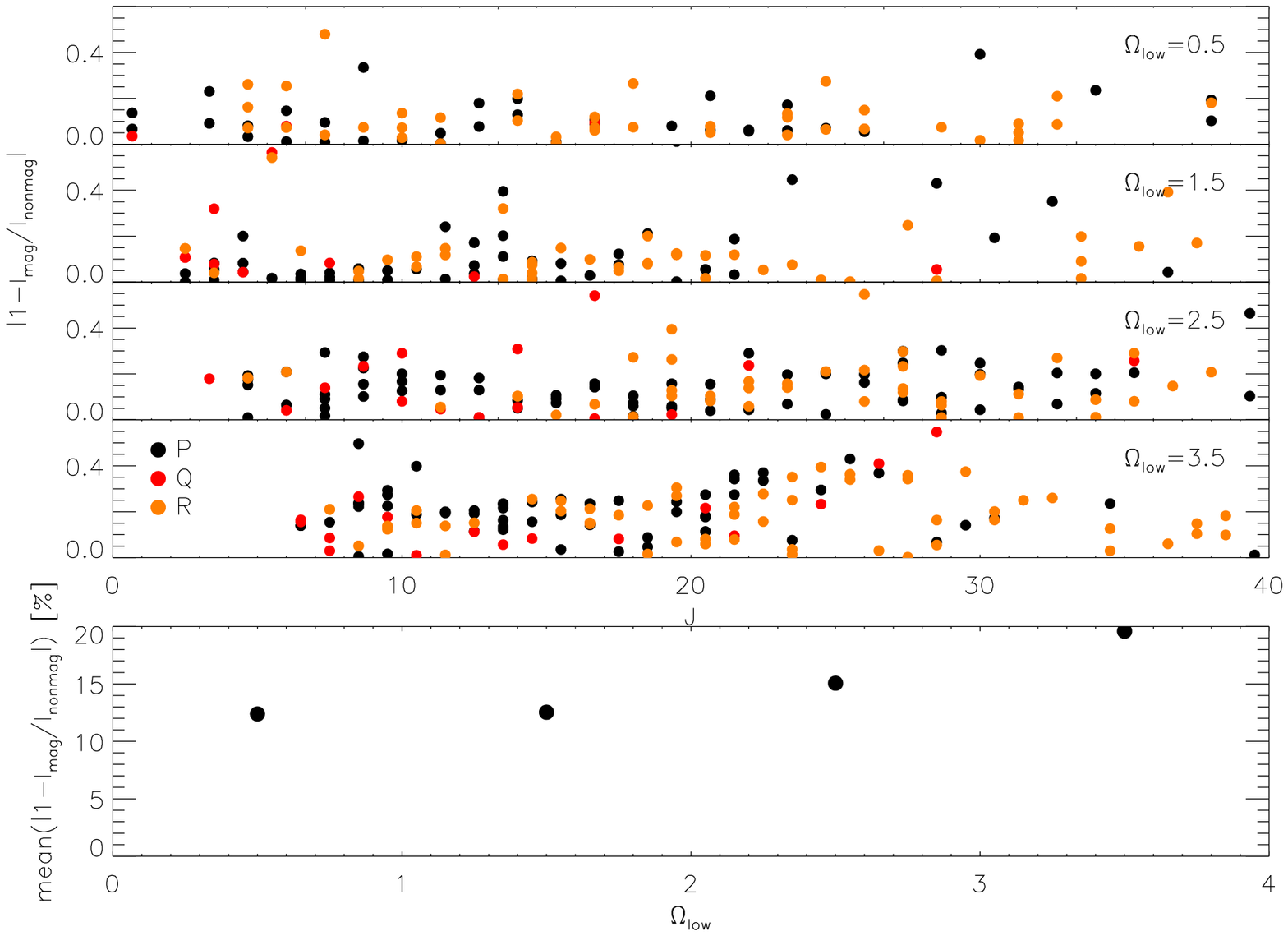}
\plotone{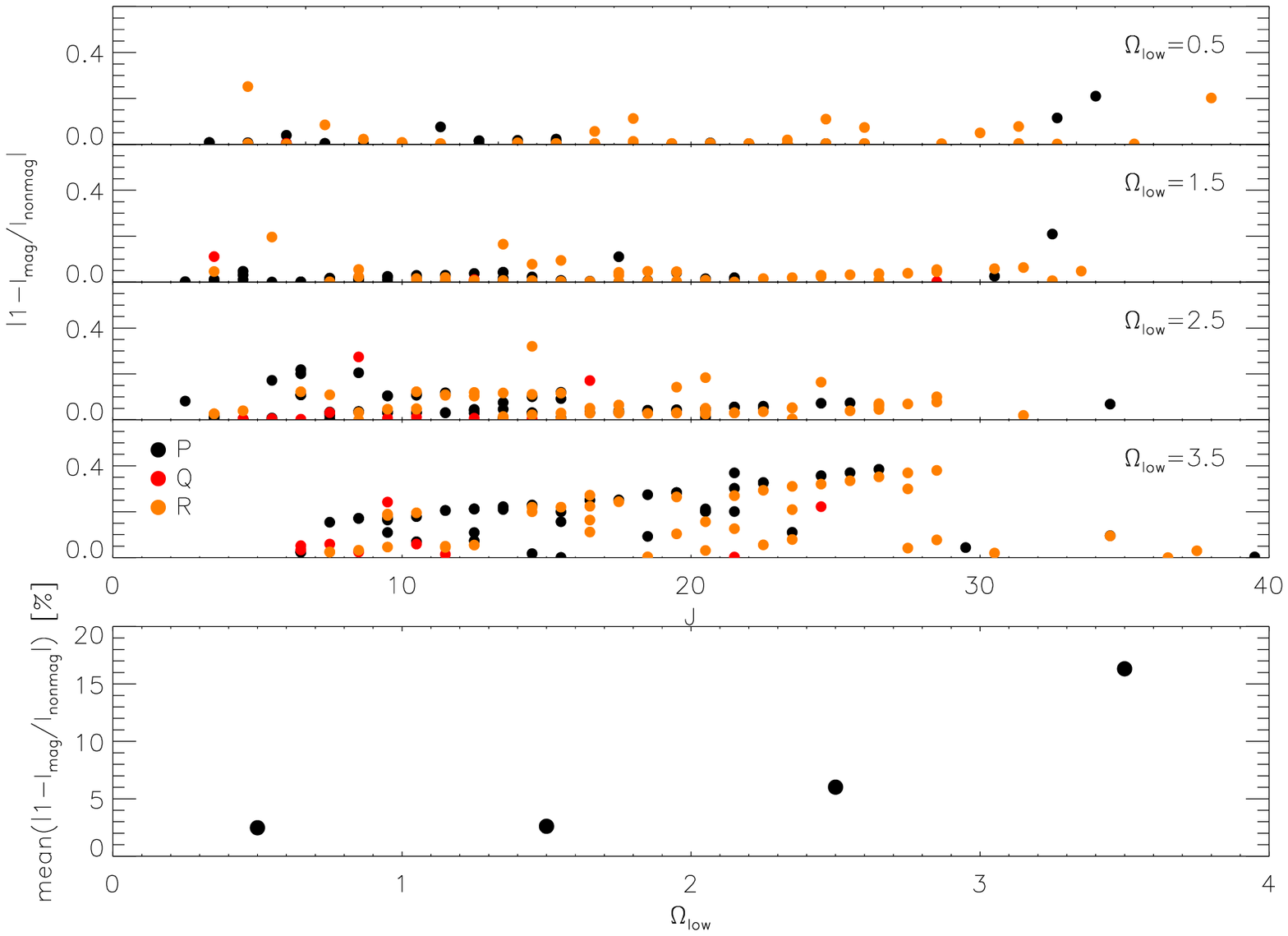}
\caption[]{Ratio between the depths of the FeH lines in the magnetic
   and non magnetic case as a function of rotational quantum number $J$. Upper plot shows the results from the
   observations, bottom plot from the computations. The lower panels
   each show the average ratio for each $\Omega$.}
\label{Zeeman3}
\end{figure}

In order to quantify the identification of magnetically sensitive
lines, we used a Gaussian fit to the FeH line profiles to measure
their depths ($I$) and widths ($\sigma$). This was done for the magnetically broadened
spectra as well as for the non magnetic ones. The ratio of the
line widths $\sigma_{mag}/\sigma_{nonmag}$ can be used to investigate if a line is
broadened by the magnetic field. The ratio $|1-I_{mag}/I_{nonmag}|$
can be used to characterise the amount of variation in the line depth.
The ratios of line widths are plotted in Fig.~\ref{Zeeman2} and the ratios of line depths are plotted in Figs.~\ref{Zeeman3}. The upper plots in these figures
are for the observations and the bottom plots for computations.
The ratios are plotted as a function of rotational
quantum number $J$ and are separated by
$\Omega$ (starting with $0.5$ at the top and ending at $3.5$ at the 
bottom in steps of $1$) since the Land\'e factor strongly depends on it
\citep{2002A&A...385..701B}. The Land\'e factor is also a function of
$J$ and different for rotational branches. Due to this, the P, Q, and
R branches are indicated by different colors. One can see that there
is no obvious dependence on $J$, which would be expected if the
splitting were pure Hund's case (a) or (b).\\
This investigation was also done for the synthetic spectra and the results are shown 
in the bottom plots of Figs.~\ref{Zeeman2} and~\ref{Zeeman3}. 
The computed spectra 
reproduce the general trends of the observations, which could be regarded as a sign
that the ansatz described above is a good approximation. 
In these figures, the average ratio
is also shown as a function of $\Omega$: the
magnetic influence is clear visible stronger for lines with high $\Omega$, in
agreement with theory. Again, the results from observations and
computations are very similar and differ only in the absolute
values. This discrepancy could be due to noise in the
observations. 

\section{Conclusion}

We conclude that the potential of FeH lines for measuring magnetic
fields is very high. Empirically, it is already possible to use them,
but the results depend on well-chosen and accurate template spectra
with known parameters. The theoretical approach is promising, but has
to be investigated further to describe the Zeeman splitting more
correctly.

\acknowledgements 

SW acknowledges funding from the GrK 1351 ``Extrasolar Planets and
their host stars''.\\
AR \& AS acknowledges research funding from the 
DFG  (RE 1664/4 -1), and DS also acknowledges funding from the DFG 
(RE 1664/7-1).\\
OK is a Royal Swedish Academy of Sciences Research Fellow supported 
by grants from the Knut and Alice Wallenberg Foundation and the 
Swedish Research Council. 

\bibliography{wende_s}

\begin{thebibliography}{}
\expandafter\ifx\csname natexlab\endcsname\relax\def\natexlab#1{#1}\fi
\expandafter\ifx\csname url\endcsname\relax
  \def\url#1{\texttt{#1}}\fi
\expandafter\ifx\csname urlprefix\endcsname\relax\def\urlprefix{URL }\fi
\providecommand{\eprint}[2][]{\url{#2}}

\bibitem[{{Afram} et~al.(2007){Afram}, {Berdyugina}, {Fluri}, {Semel},
  {Bianda}, \& {Ramelli}}]{2007A&A...473L...1A}
{Afram}, N., {Berdyugina}, S.~V., {Fluri}, D.~M., {Semel}, M., {Bianda}, M., \&
  {Ramelli}, R. 2007, \aap, 473, L1. \eprint{0708.0298}

\bibitem[{{Afram} et~al.(2008){Afram}, {Berdyugina}, {Fluri}, {Solanki}, \&
  {Lagg}}]{2008A&A...482..387A}
{Afram}, N., {Berdyugina}, S.~V., {Fluri}, D.~M., {Solanki}, S.~K., \& {Lagg},
  A. 2008, \aap, 482, 387

\bibitem[{{Berdyugina} \& {Solanki}(2002)}]{2002A&A...385..701B}
{Berdyugina}, S.~V., \& {Solanki}, S.~K. 2002, \aap, 385, 701

\bibitem[{{Berdyugina} et~al.(2003){Berdyugina}, {Solanki}, \&
  {Frutiger}}]{2003A&A...412..513B}
{Berdyugina}, S.~V., {Solanki}, S.~K., \& {Frutiger}, C. 2003, \aap, 412, 513

\bibitem[{{Kochukhov}(2007)}]{2007pms..conf..109K}
{Kochukhov}, O.~P. 2007, in Physics of Magnetic Stars, 109.
  \eprint{arXiv:astro-ph/0701084}

\bibitem[{{Leroy}(2004)}]{Bleroy}
{Leroy}, B. 2004

\bibitem[{{Reid} \& {Hawley}(2005)}]{2005nlds.book.....R}
{Reid}, I.~N., \& {Hawley}, S.~L. 2005, {New light on dark stars : red dwarfs,
  low-mass stars, brown dwarfs}

\bibitem[{{Reiners} \& {Basri}(2006)}]{2006ApJ...644..497R}
{Reiners}, A., \& {Basri}, G. 2006, ApJ, 644, 497.
  \eprint{arXiv:astro-ph/0602221}

\bibitem[{{Reiners} \& {Basri}(2007)}]{2007ApJ...656.1121R}
--- 2007, ApJ, 656, 1121. \eprint{arXiv:astro-ph/0610365}

\bibitem[{{Shulyak} et~al.(2010){Shulyak}, {Reiners}, {Wende}, {Kochukhov},
  {Piskunov}, \& {Seifahrt}}]{2010A&A...523A..37S}
{Shulyak}, D., {Reiners}, A., {Wende}, S., {Kochukhov}, O., {Piskunov}, N., \&
  {Seifahrt}, A. 2010, \aap, 523, A37+. \eprint{1008.2512}

\bibitem[{{Wende} et~al.(2010){Wende}, {Reiners}, {Seifahrt}, \&
  {Bernath}}]{2010A&A...523A..58W}
{Wende}, S., {Reiners}, A., {Seifahrt}, A., \& {Bernath}, P.~F. 2010, \aap,
  523, A58+. \eprint{1007.4116}

\end{thebibliography}

\end{document}